\documentclass[10pt,journal]{IEEEtran}

\usepackage{cite}
\usepackage{graphicx}
\usepackage{amsthm}
\usepackage{mathtools}
\usepackage{amsmath}
\usepackage{amssymb}
\usepackage{dsfont}
\usepackage{mathrsfs}
\usepackage{multirow}
\usepackage{booktabs}
\usepackage{algpseudocode}
\usepackage{algorithm}
\usepackage{algorithmicx}
\usepackage{setspace}
\usepackage{xspace}
\newcommand*\Let[2]{\State #1 $\gets$ #2}
\algrenewcommand\algorithmicrequire{\textbf{Input:}}
\algrenewcommand\algorithmicensure{\textbf{Output:}}
\newtheorem{definition}{Definition}

\begin{document}
\title{High-efficiency Blockchain-based\\Supply Chain Traceability}
\author{Hanqing Wu, Shan Jiang, Jiannong Cao, \textit{Fellow, IEEE}
    \thanks{Hanqing Wu, Shan Jiang, and Jiannong Cao were with the Department of Computing, The Hong Kong Polytechnic University.}
    \thanks{Emails: hanqing.91.wu@connect.polyu.hk, cs-shan.jiang@polyu.edu.hk, jiannong.cao@polyu.edu.hk.}
    \thanks{Corresponding author: Shan Jiang.}
}
\maketitle
\begin{abstract}
Supply chain traceability refers to product tracking from the source to customers, demanding transparency, authenticity, and high efficiency. In recent years, blockchain has been widely adopted in supply chain traceability to provide transparency and authenticity, while the efficiency issue is understudied. In practice, as the numerous product records accumulate, the time- and storage- efficiencies will decrease remarkably. To the best of our knowledge, this paper is the first work studying the efficiency issue in blockchain-based supply chain traceability. Compared to the traditional method, which searches the records stored in a single chunk sequentially, we replicate the records in multiple chunks and employ parallel search to boost the time efficiency. However, allocating the record searching primitives to the chunks with maximized parallelization ratio is challenging. To this end, we model the records and chunks as a bipartite graph and solve the allocation problem using a maximum matching algorithm. The experimental results indicate that the time overhead can be reduced by up to $85.1$\% with affordable storage overhead.
\end{abstract}

\section{Introduction}
In $2019$, the global supply chain market value surpassed $14.6$ trillion US dollars, having increased at a compound annual growth rate of $10.8$\% since $2015$ \cite{beamon1998supply}. The supply chain plays a vital role in the global economy. Supply chain management, which refers to the flow management of goods and services, including all the processes that transform raw materials into final products along the supply chain, is essential for boosting customer services, reducing operation costs, improving financial positions, etc. \cite{mentzer2001defining}

Among supply chain management services, traceability is essential because it allows product tracking from the sources to end consumers \cite{lee2008rfid}. Supply chain traceability provides opportunities to enhance the supply chain efficiencies, meet the regulatory requirements, and, most importantly, to story-tell the consumers about the provenance and journey of products. Regarding the products whose safety is critical, e.g., food and pharmaceuticals, supply chain traceability is critical and has been pursued for decades by the industries \cite{kelepouris2007rfid}.

Despite its importance, supply chain traceability is challenging primarily because of its undue reliance on the collaboration of multiple stakeholders who are not motivated to collaborate. Moreover, there is a lack of mechanisms to define the minimum amount of data required from the stakeholders to achieve supply chain traceability. In existing studies, the researchers focused on modeling supply chain traceability, especially the data among different stakeholders \cite{hu2013modeling,bechini2008patterns}. The incentives of collaboration are understudied, letting alone the safety and quality of the tracing process \cite{aung2014traceability}.

In recent years, blockchain has been regarded as a promising solution for supply chain traceability because of the distinctive features of immutability, transparency, auditability, and native support of incentivization \cite{gurtu2019potential,chang2020blockchain}. Generally, a blockchain is an append-only list of blocks, each containing a set of transactions, maintained by a decentralized peer-to-peer network \cite{jiang2018blochie}. The product records stored on the blockchain are publicly available and cannot be modified, making the stored information reliable. The auditability makes it possible to track product information on a blockchain. Furthermore, blockchain natively embeds tokens to incentivize collaboration among supply chain stakeholders. To summarize, blockchain empowers supply chain traceability with high reliability, auditability, and incentives for collaboration \cite{WambaQ20,queiroz2019blockchain,wu2019data}.

Besides the applications of blockchain-based supply chain traceability in big enterprises such as IBM and Walmart \cite{kamath2018food}, blockchain solutions for supply chain traceability are also extensively studied in academia. On the one hand, the concept of blockchain-based supply chain traceability and the corresponding system design are discussed in many research works \cite{apte2016will,hackius2017blockchain,blossey2019blockchain,saberi2019blockchain,cole2019blockchain}.
On the other hand, the researchers find blockchain technology can be used together with other technologies, such as the Internet of Things (IoT), to provide the traceability service \cite{tian2016agri,sidorov2019ultralightweight,mondal2019blockchain,jangirala2019designing,jiang2021fairness}.
However, all these works in industry and academia focus on the design of the traceability system while leaving the efficiency issue alone. In practice, the time- and storage- efficiencies are significantly affected by the considerable and increasing number of product records generated by the ubiquitous IoT devices.

To the best of our knowledge, this paper is the first work studying high-efficiency blockchain-based supply chain traceability. In particular, we demonstrate the system architecture of a blockchain-based supply chain and model the product records as a directed acyclic graph. To this end, the traceability problem is defined as a graph searching problem over the blockchain. To address the problem, we propose replicating the product records in multiple chunks in a database and developing a novel parallel search algorithm based on the maximum matching algorithm to improve searching efficiency significantly. The fundamental principle of efficiency improvement lies in sacrificing storage overhead to reduce time overhead. The key technical depth lies in the matching-based parallel search algorithm.

The main contributions of this work are as follows:
\begin{itemize}
    \item To the best of our knowledge, we are the first to study and formally model the high-efficiency issue in blockchain-based supply chain traceability.
    \item We propose a novel parallel search algorithm based on the maximum matching algorithm, which significantly boosts product tracking efficiency.
    \item We conduct extensive experiments on the proposed algorithm, which indicates up to $85.1$\% time reduction for product tracking.
\end{itemize}

The rest of this paper is organized as follows. Sec.~\ref{sec:related-work} introduces the related work of this work. Sec.~\ref{sec:preliminaries} provides the preliminaries of the problem. In Sec.~\ref{sec:system-problem}, we explain the system model and formally define the problem of high-efficiency blockchain-based supply chain traceability. Sec.~\ref{sec:algorithm} gives the traditional approach and the proposed algorithm for solving the traceability problem. Sec.~\ref{sec:experiments} demonstrates the experimental results. Finally, Sec.~\ref{sec:conclusion} concludes this work and discusses the future directions.

\section{Related Work}\label{sec:related-work}
In this section, we survey the related work about high-efficiency blockchain-based supply chain traceability, i.e., supply chain traceability and searching over blockchain, and articulate the motivations and novelty of this work.

\subsection{Supply Chain Traceability}
The research on supply chain traceability can be roughly divided into two categories, i.e., unified data representation methods for various stakeholders along the supply chain, and digital technologies to facilitate reliable and ubiquitous information storage.

A large number of stakeholders along the supply chain have their own data management systems with diversified data formats. Supply chain traceability needs to retrieve the data from the stakeholders, and a unified data representation method is demanded. The unified data representation methods for supply chain information have been studied for years. In \cite{bechini2008patterns}, Bechini et al. investigate the issues for supply chain traceability, introduce a traceability data model and a set of suitable patterns, discuss the suitable technological standards to define, register, and enable business collaborations, and implement a real-world system for food supply chain traceability.
In \cite{hu2013modeling}, Hu et al. propose a Unified Modeling Language (UML) model for traceability along with a set of suitable patterns, develop a series of UML class diagrams to conceive a method for modeling the product, process, and quality information along the supply chain, and conduct a case study on vegetable supply chain traceability.

Regarding digital technologies for supply chain traceability, radio-frequency identification (RFID) and blockchain are representative. In particular, RFID is a sensing technology that helps to collect the data along supply chains ubiquitously, while blockchain is a distributed ledger technology to provide secure and reliable data storage services.

The usage of RFID in supply chain traceability can be traced back to as early as $2003$ \cite{karkkainen2003increasing}, at which time Karkkainen proposed to develop an RFID-based data capture system to solve the problems associated with the logistics of short shelf-life products. In $2007$, RFID was widely recognized as a promising technology for supply chain traceability \cite{attaran2007rfid,kelepouris2007rfid} because the passive RFID tags on the products are cheap, do not need to be within the line of sight of the RFID reader (compared with barcodes), and do not need batteries (compared with other sensors).
Later, there are also surveys about RFID-enabled supply chain traceability \cite{sarac2010literature,wu2011rfid,costa2013review}.

The potential of using blockchain technology for supply chain traceability was investigated by Tian in $2016$ for the first time \cite{tian2016agri}, in which a traceability system was designed for agri-food supply chains combining RFID and blockchain technologies. Although the work is a pioneer, it is conceptual without real-world deployment. We see that the product information recorded on a blockchain is immutable, i.e., it cannot be modified once stored, making the traceability results reliable. Similar works include \cite{abeyratne2016blockchain,tian2017supply,biswas2017blockchain,caro2018blockchain,francisco2018supply,wang2020blockchain} in the supply chains of construction, wine, etc., some of which are implemented in real-world settings.

\subsection{Searching over Blockchain}
Blockchain-based supply chain traceability requires blockchain data to be searched given a product item. We present the related work about searching over blockchain in this subsection. In particular, searching over blockchain refers to the process that the users (with no local storage) request blockchain full nodes (with full storage) to search data on a blockchain, in which the search requests can be keyword search, range query, etc. In literature, integrity, privacy, and efficiency are the three concerned performance metrics of searching over blockchain, in which integrity means whether the search results are sound and complete, privacy means whether data leakage happens during searching, and efficiency means the time and communication overhead.

The naive procedure of searching over the blockchain is as follows. First, the user sends a searching request to a blockchain full node. Then, the full node proceeds with the request by scanning the data on the blockchain block by block and transaction by transaction, and recording all the data satisfying the searching request. Finally, the full node returns the search result. As we can see, the integrity of the search result cannot be guaranteed, the privacy can be disclosed because of the raw data on the blockchain, and the efficiency is low because scanning transactions one by one takes a long time. The research community has been developing solutions to improve integrity, privacy, and efficiency.

Smart contracts and verifiable computation are the two approaches to guarantee searching integrity. The basic idea of the smart contract is to send the searching requests to all the blockchain nodes rather than a single one. The incentive mechanism of blockchain will motivate the majority of the blockchain nodes to return sound and complete search results, which guarantees integrity. The advantage of using smart contracts is that the method is general and can be easily adapted to all kinds of data and queries. However, the drawback lies in the high cost of executing smart contracts.
In terms of verifiable computation, the search result returned to the user will be accompanied by proof for integrity verification. Using verifiable computation can fine-tune the efficiency by designing subtle data structure \cite{xu2019vchain,zhang2019gem,hao2020outsourced,zhang2021authenticated,guo2020blockchain,dai2020lvq}. In contrast, the disadvantage is that there is no general data structure for all types of data and queries.

Searchable encryption is the major approach for privacy preservation during searching over blockchains. The data, queries, and search results are encrypted compared with the naive search approach. The research community has been developing efficient searchable encryption scheme for various types of data and queries \cite{hu2018searching,jiang2019privacy,guo2020verifiable,ding2020enabling,chen2019blockchain}.

To summarize, the existing studies about blockchain-based supply chain traceability mainly focus on the system design while leaving the efficiency issue alone. When we reduce blockchain-based supply chain traceability to a problem of searching over blockchain, we find that the reduced graph searching problem on blockchains is new.

\section{Preliminaries}\label{sec:preliminaries}
In this section, we introduce the preliminary knowledge about blockchain data structure and maximum matching algorithms for bipartite graphs. Note that the maximum matching algorithm is a necessary component for maximizing the parallelization ratio in Sec.~\ref{sec:algorithm}.

\subsection{Blockchain Data Structure}

\begin{figure}[!ht]
\centering
\includegraphics[width=\linewidth]{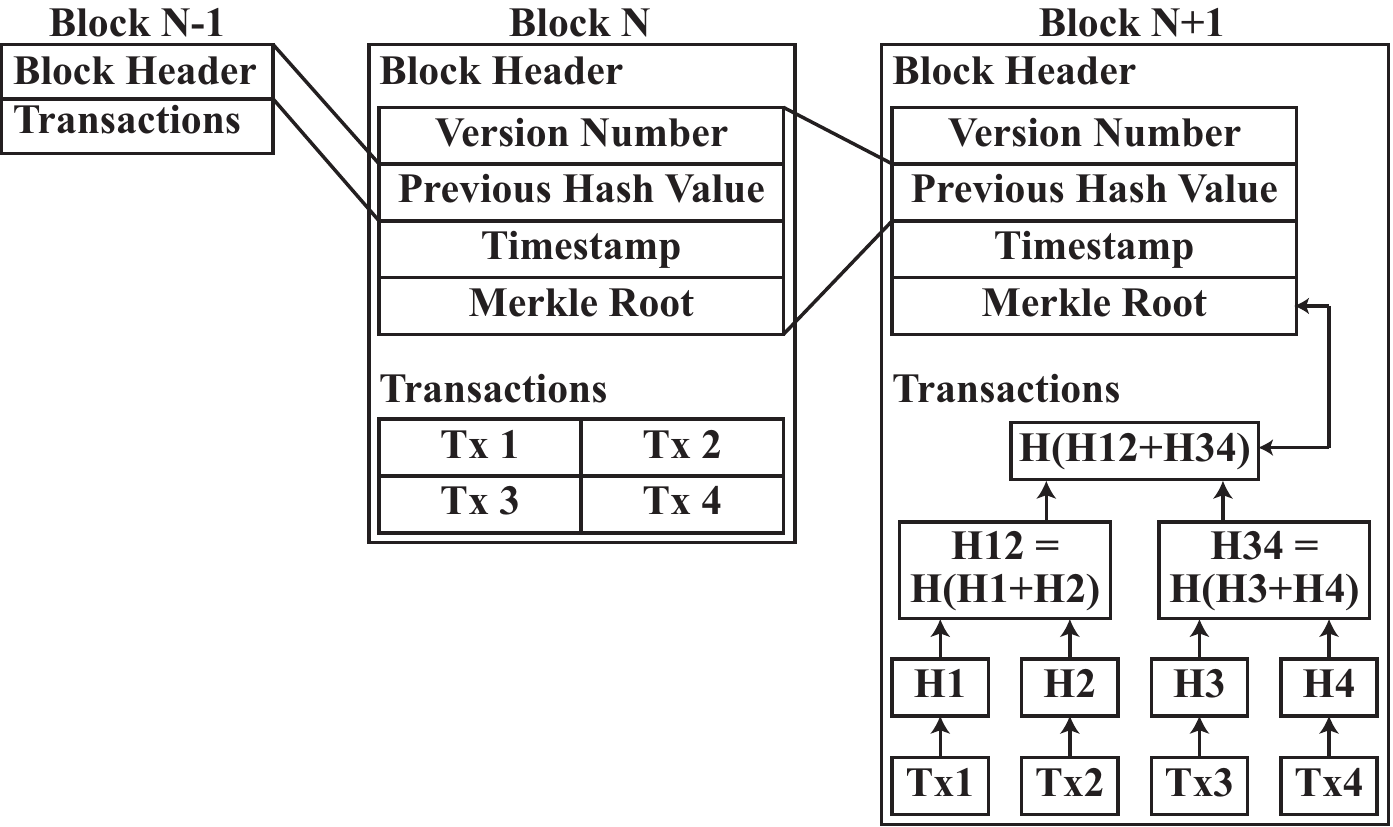}
\caption{Structure of a typical blockchain. The blocks are linked into a chain using cryptographic hash values.}\label{fig:blockchain}
\end{figure}

A blockchain is an append-only list of blocks linked by cryptographic values, in which each block contains a set of transactions, maintained by a decentralized peer-to-peer network \cite{wang2019survey}. Fig.~\ref{fig:blockchain} depicts the structure of blockchain with description. Specifically, a single valid block consists of a block header and a list of transactions. The following fields briefly document the block details:

\begin{itemize}
    \item \textit{Block Header} provides the important information inside the block. It includes the Version Number, Previous Hash Value, Timestamp, Merkle root, etc. Each block header is hashed, unique, and cryptographically secured, supporting the immutability property of blockchains. For example, in Bitcoin \cite{cusumano2014bitcoin}, ``target difficulty'' and ``nonce'' are included as part of the Proof of Work (PoW) consensus algorithm used when mining.
    \item \textit{Version Number} indicates which version of block validation rules to follow. If the block version number differs from other blocks, it means this block is running on a different chain, commonly known as a hard fork.
    \item \textit{Previous Hash Value} is a byte field containing the hash of the previous block header, serving as a pointer to the previous block. Such a field ensures that no previous block can be modified without changing the current block header, making the whole blockchain difficult and even impossible to be modified.
    \item \textit{Timestamp} is the time of generating this block which is more commonly known as the time when the miner started hashing the current block header. It can be used to calculate the average block propagation time.
    \item \textit{Merkle root} is derived from the hashes of all transactions included in the current block. It is a tamper resistance measure that those transactions cannot be modified without changing the Merkle Root value, furthermore, the entire header. Merkle root is also a fast and efficient way to verify the data. In Fig.~\ref{fig:blockchain}, the Merkle root of block $N+1$ is computed as the hierarchical hash results upon the transactions inside.
    \item \textit{Transactions} contains the transactions confirmed by the blockchain network and packed in the block. For example, in Bitcoin \cite{cusumano2014bitcoin}, a typical transaction represents the money transfer of two or more parties.
\end{itemize}

\subsection{Maximum Matching}
In graph theory, a matching in an undirected graph is a set of edges without common vertices. The maximum matching problem is to find a matching that uses as many edges as possible given an undirected graph.

A bipartite graph is a graph whose vertices can be divided into two disjoint and independent sets $U$ and $V$ such that every edge connects a vertex in $U$ and a vertex in $V$. The maximum matching problem on a bipartite graph is well studied and can be solved efficiently (in polynomial time) using the Hungarian algorithm \cite{kuhn1955hungarian}, Ford-Fulkerson algorithm, etc.

\begin{figure*}[!ht]
\centering
    \includegraphics[width=.8\textwidth]{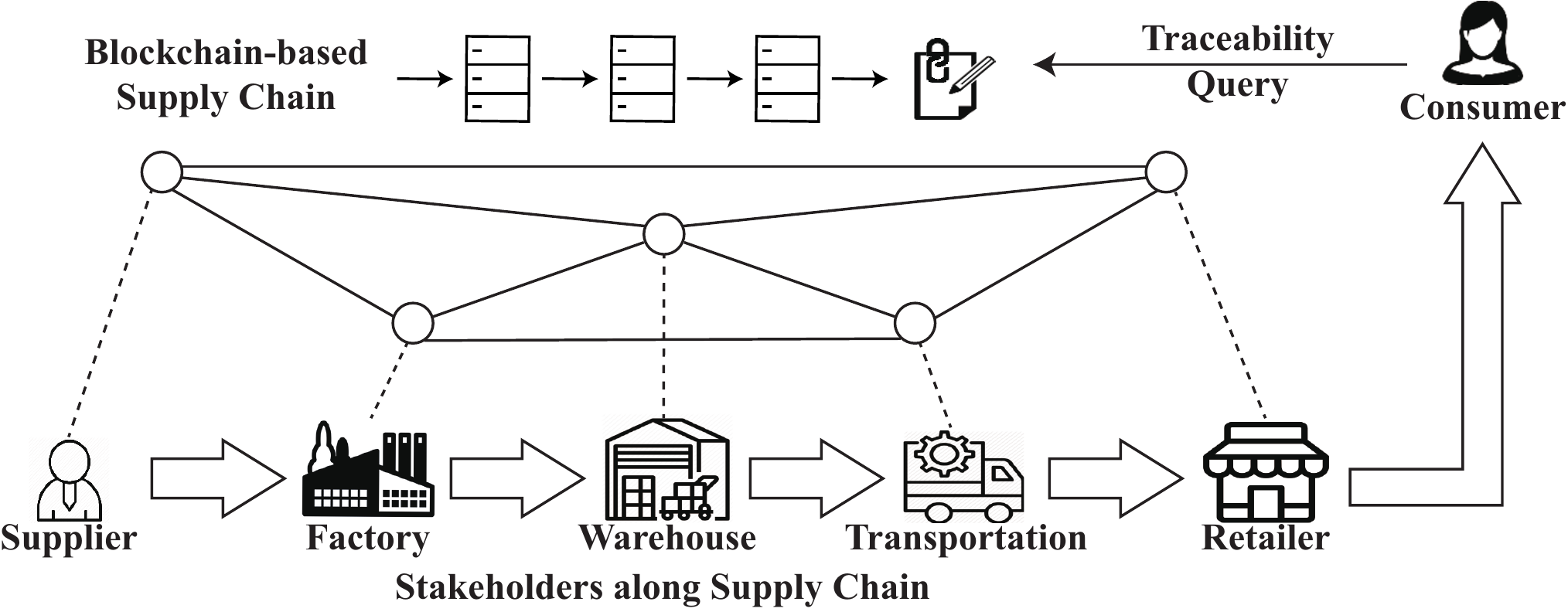}
    \caption{System model of the blockchain-based supply chain. There are three layers: the bottom layer contains the stakeholders along the supply chain; the middle layer is the blockchain network maintained by the supply chain stakeholders; the top layer is the blockchain data and smart contract, providing traceability services to the consumers.}
    \label{fig:system-model}
\end{figure*}

\section{System Model and Problem Definition}\label{sec:system-problem}
This section first gives the system model of the blockchain-based supply chain and then formally defines the problem of high-efficiency blockchain-based supply chain traceability.

\subsection{System Model}\label{subsec:system}

Fig.~\ref{fig:system-model} elaborates on the system model of the blockchain-based supply chain. In particular, the stakeholders along the supply chain, e.g., raw material suppliers, factories, warehouses, transportation companies, and retailers, form a peer-to-peer network and maintain a permissioned blockchain. The regulatory authorities can also join and expand the blockchain network. Our system prefers permissioned blockchain to the public one because the nodes not hosted by the supply chain stakeholders should be forbidden from joining the blockchain network. Note that each stakeholder may contribute a set of blockchain nodes, and the whole blockchain network will be of large scale. In our system, the stakeholders will upload the product information to the blockchain motivated by the following reasons. First, transparent product information on the blockchain will strengthen the consumers' confidence in the products. Second, the inter-related information helps improve the efficiency of supply chain management. Finally, the product information will better meet the frequent regulation requests. Note that the blockchain system can only guarantee that the information cannot be tampered with once stored. If a stakeholder provides incorrect records, the blockchain can provide non-tamperable and permanent proof of the incorrectness. In terms of the end consumers, they will enjoy the services provided by the supply chain, as well as query the product tracking information through the blockchain.

The product information recorded on the blockchain will contain at least the following fields:
\begin{itemize}
    \item \textsc{Time}: the timestamp when the record is submitted.
    \item \textsc{Location}: the location when the record is submitted.
    \item \textsc{Publisher}: the one who submitted the record.
    \item \textsc{SrcItems}: the unique identifiers of the source (original) food items.
    \item \textsc{DesItems}: the unique identifiers of the destination (result) food items.
\end{itemize}

The fields of \textsc{SrcItems} and \textsc{DesItems} indicate the relationships among the product. That is, the products in \textsc{SrcItems} are the raw materials of the ones in \textsc{DesItems}. When talking about supply chain traceability, the products in \textsc{SrcItems} should be output if any product in \textsc{DesItems} is set as the input.

\subsection{Problem Definition}\label{subsec:problem}
In this section, we define the problem of high-efficiency traceability formally.

Generally speaking, the function of blockchain is to serialize a set of transactions to an ordered list.

\begin{definition}
A \textbf{blockchain} $\mathcal{B} = (t_1, t_2, \cdots)$ is defined to be an append-only list of transactions, in which $t_i$s are transactions.
\end{definition}

The transactions $t_i$s in the blockchain are totally ordered, which means $t_j$ is confirmed after $t_i$ for sure if $i<j$.

\begin{definition}
In a blockchain, a \textbf{transaction} $t_i = (id_i, \mathcal{P}_i)$ is defined to be a tuple of identifier and direct predecessors, in which $id_i$ is the identifier while $\mathcal{P}_i$ is the set of identifiers of direct predecessor transactions.
\end{definition}

In the context of traceability, the predecessor means the relationship of dependency, e.g., a bag of potato chips is made from a package of potatoes.
Note that for a given transaction $t_i$, the predecessors in $\mathcal{P}_i$ must be already there in the blockchain, e.g., the transaction of potatoes must appear before the transaction of potato chips in the blockchain.
Formally speaking, if $id_j \in \mathcal{P}_i$, then we can infer that $j < i$.

For better understanding, the relationship among the transactions can be represented as a direct acyclic graph (DAG). The construction of the DAG given a blockchain is as follows:
\begin{itemize}
    \item for each transaction $t_i$, add a vertex $v_i$; and
    \item for each transaction $t_i$ and each identifier $id_j \in \mathcal{P}_i$, add a directed edge from $v_j$ to $v_i$.
\end{itemize}

\begin{figure}[!ht]
\centering
\includegraphics[width=.6\linewidth]{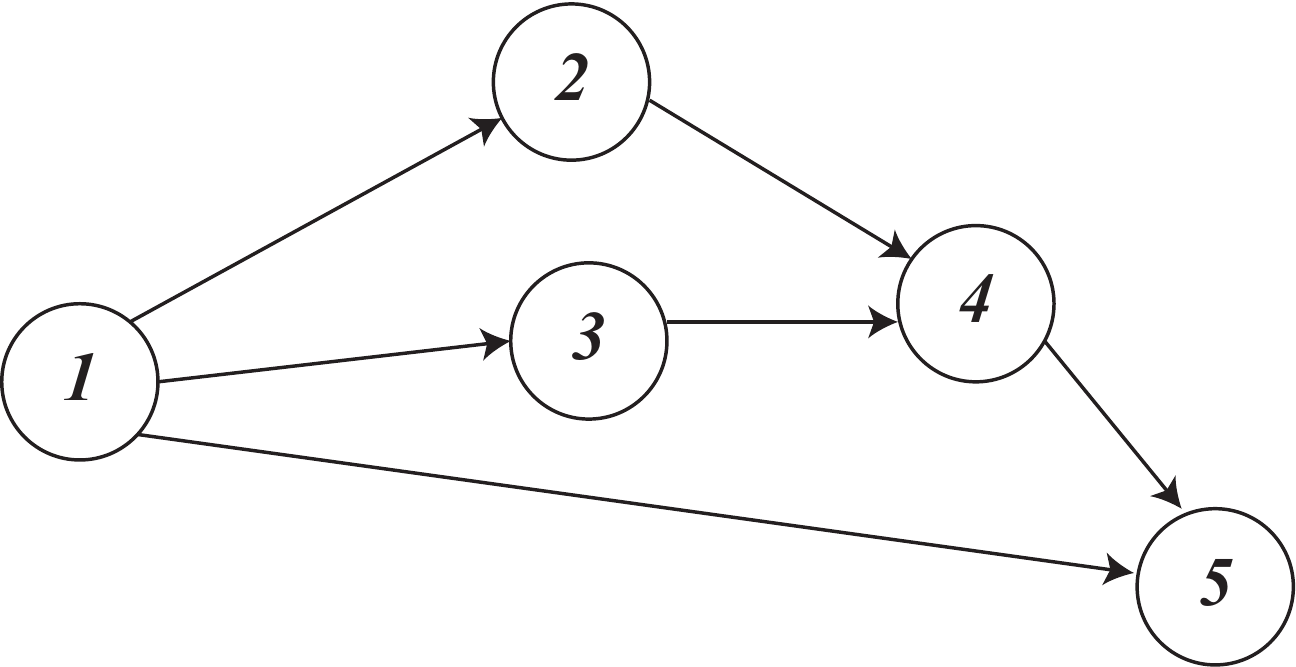}
\caption{Example DAG based on the transactions in Tab.~\ref{tab:example-transactions}. In the DAG, the nodes represent the transactions, while the edges represent the predecessor relationship. For example, node $1$ has no incoming edge because transaction $1$ has no predecessor.}\label{fig:example-dag}
\end{figure}

An example set of transactions and its corresponding DAG are shown in Tab.~\ref{tab:example-transactions} and Fig.~\ref{fig:example-dag}, respectively.
In the example, the transactions with identifiers $1$ and $4$ are the direct predecessors of the transaction $5$.
Meanwhile, transaction $3$ is a predecessor (indirect) of $5$ as in Fig.~\ref{fig:example-dag}.
In this work, we define \textit{traceability} as a function to find all the predecessors (both direct and indirect) of a given transaction in a given blockchain.
The formal definitions of \textit{direct predecessor}, \textit{predecessor} and \textit{traceability} are given are follows.

\begin{table}[!ht]
    \centering
    \caption{Example Transactions in Blockchain}\label{tab:example-transactions}
    \begin{tabular}{cc}
    \toprule
    Identifier & Direct Predecessors \\
    \midrule
    1 & $\emptyset$\\
    2 & $\{1\}$ \\
    3 & $\{1\}$ \\
    4 & $\{2,3\}$ \\
    5 & $\{1,4\}$ \\
    \bottomrule
    \end{tabular}
\end{table}

\begin{definition}
A transaction $t_i$ is defined to be a \textbf{direct predecessor} of another transaction $t_j$ if $id_i \in \mathcal{P}_j$.
\end{definition}

\begin{definition}
A transaction $t_i$ is defined to be a \textbf{predecessor} of another transaction $t_j$ if there is a list of transactions $t_{k_1}, t_{k_2}, \cdots, t_{k_l}$ such that $id_i \in \mathcal{P}_{k_1}$, $id_{k_1} \in \mathcal{P}_{k_2}$, $\cdots$, $id_{k_{l-1}} \in \mathcal{P}_{k_l}$, and $id_{k_l} \in \mathcal{P}_{j}$.
\end{definition}

In a blockchain, we assume that there is a function called \textsc{GetPredecessors}, which takes a transaction identifier as input and outputs all the direct predecessors of the input transaction. We assume that \textsc{GetPredecessors} takes $t(n)$ time in which $n$ is the number of transactions in the blockchain. Note that the expression $t(n)$ depends on the implementation of \textsc{GetPredecessors}. For example, if \textsc{GetPredecessors} is implemented using a binary search tree, then $t(n) = O(\log n)$.

\begin{definition}
Problem \textbf{Traceability}: given a blockchain and a transaction identifier $id_i$, output the identifiers of all the predecessors of $t_i$.
\end{definition}

\begin{table}[!ht]
    \centering
    \caption{Example Input and Output of Traceability}\label{tab:example-traceability}
    \begin{tabular}{cc}
    \toprule
    Input & Output \\
    \midrule
    1 & $\emptyset$\\
    2 & $\{1\}$ \\
    3 & $\{1\}$ \\
    4 & $\{1,2,3\}$ \\
    5 & $\{1,2,3,4\}$\\
    \bottomrule
    \end{tabular}
\end{table}

Following the definition of \textit{traceability}, if the input is transaction $4$, then the output should be transactions $1$, $2$, and $3$. Other examples of input and output can be found in Tab.~\ref{tab:example-traceability}.

\section{Proposed Algorithm \& Analysis}\label{sec:algorithm}
In this section, we present the traditional algorithm and the proposed algorithm for solving the problem \textit{traceability}.

\subsection{Traditional Approach}\label{TraAppr}
The naive approach to solving \textit{traceability} is breadth-first search (BFS) as shown in Algo.~\ref{alg:bfs}.

\begin{algorithm}
\caption{Breadth-first search algorithm to solving the problem \textit{traceability}} \label{alg:bfs}
\algorithmicrequire{ $\mathcal{B} = (t_1, t_2, \cdots, t_n)$: a blockchain of $n$ transactions; $id$: identifier of a transaction}\\
\algorithmicensure{ $\mathcal{AP}$: all the predecessors of $t_i$}
\begin{algorithmic}[1]
\Let{$\mathcal{AP}$}{$\emptyset$}
\Let{$Q$}{a first-in-first-out queue with a single element $id$}
\While{$Q$ is not empty}
    \Let{$u$}{\textsc{PopQueue}($Q$)}
    \Let{$\mathcal{P}_u$}{\textsc{GetPredecessors}($u$)}
    \For{each $v\in \mathcal{P}_u$}
        \If{$v \notin \mathcal{AP}$}
            \State{\textsc{PushQueue}($Q$, $v$)}
            \Let{$\mathcal{AP}$}{$\mathcal{AP} \cup \{v\}$}
        \EndIf
    \EndFor
\EndWhile
\State\Return{$\mathcal{AP}$}
\end{algorithmic}
\end{algorithm}

In this straightforward solution, searching is time-consuming when repeatedly accessing the index and block. In particular, we have a set  $\mathcal{AP}$, which is the expected output of the given transaction $id_i$. The $\mathcal{AP}$ is empty at the beginning. A first-in-first-out queue, $Q$, is created to hold all the elements that need to be processed. While the $Q$ is not empty, we pick out one element $u$ at a time, \textsc{PopQueue} this element $u$ from the queue. The function \textsc{GetPredecessors} is called to get the direct predecessor or predecessors of $u$ and temporarily cached at $\mathcal{P}_u$. 

For each element $\mathcal{P}_u$, if it is not in $\mathcal{AP}$, which means it is a new element, $\mathcal{P}_u$ will be \textsc{PushQueue} to the queue $Q$. At the same time, we update $\mathcal{AP}$ with the new element $\mathcal{P}_u$. If $\mathcal{P}_u$ is already in $\mathcal{AP}$, which means the element has already been processed, no further operation will be needed. This procedure stops when the queue $Q$ is empty, indicating that the entire blockchain has been gone through. This procedure processes all the elements linearly, one element at a time. Although the breadth-first search-based solution achieves the objective of \textit{traceability}, the efficiency is quite low because only one element can be processed at a time.

\subsection{Proposed Solution}\label{subsec:prosol}

The critical drawback of the traditional approach lies in the frequent operations of \textsc{GetPredecessors} of high time overhead. To improve the time efficiency, we gain the insight that the operations of \textsc{GetPredecessors} can be parallelized when tracing the transactions. In particular, we use $\alpha$ chunks to store the transactions, which can be accessed in parallel. Each transaction is replicated for $\beta - 1$ times in the chunks to enhance the degree of parallelization.

To store the transactions into $\alpha$ chunks, we need to find a way to evenly distribute all transactions into chunks without causing an imbalance in the chunk storage. Here we propose Algo.~\ref{alg:tx-allocation}, a transaction allocation mechanism. We directly modulo each transaction identifier with $\alpha$ and store this transaction pair $(id_i, \mathcal{P}_i)$ into the corresponding chunk.

\begin{algorithm}
\caption{Transaction allocation} \label{alg:tx-allocation}
\algorithmicrequire{ $id_i$: identifier of the new transaction; $\mathcal{P}_i$: the set of direct predecessors of the new transaction}\\
\algorithmicensure{ The allocation scheme of the new transaction}
\begin{algorithmic}[1]
\For{$i \gets 0$ \textbf{to} $\beta - 1$}
    \State{Store $(id_i, \mathcal{P}_i)$ in $\mathcal{CK}_{id_i \bmod{\alpha}}$}
\EndFor
\end{algorithmic}
\end{algorithm}

With transactions evenly allocated to $\alpha$ chunks, we propose the parallelized search algorithm as shown in Algo.~\ref{alg:parallel-trace}. To recap, with given transaction identifier $id_i$, we aim at getting all the predecessors $\mathcal{AP}$ of $id_i$.

The $\mathcal{AP}$ is empty at first, the same with Algo.~\ref{alg:bfs}. Then, a set $S$ is created to hold the elements that need to be processed. While the $S$ is not empty, a scheduling algorithm \textsc{Schedule} will be called to generate optimal transaction-chunk pairs $\mathcal{R}$ from the $S$. This $\mathcal{R}$ contains a list of transaction-chunk pairs which has no conflict with each other. The primary purpose is to process different elements stored in different chunks simultaneously since processing them one by one essentially hinders the searching efficiency, as discussed in Sec.~\ref{TraAppr}. The details of the procedure \textsc{Schedule} is explained in Algo.~\ref{alg:schedule}.

For each pair $(id_i, \mathcal{CK}_i)$ from $\mathcal{R}$, a new thread is forked exclusively to handle the pair. \textsc{GetPredecessors} will be called to get the direct predecessors of $id_i$, and $id_i$ will be removed from $S$. We wait until all the threads, forked from each pair, terminate. For the results returned from each thread, if the element is not in $\mathcal{AP}$, which means it is a new element, the element is added to both $S$ and $\mathcal{AP}$. Otherwise, it is a processed element that needs no further operation. Please note that the condition is whether $id$ is not in $\mathcal{AP}$ since the $\mathcal{AP}$ is the expected result set while the $S$ is a set with elements awaiting to be processed. The $\mathcal{AP}$ is the superset of $S$. For example, a processed element will be in $\mathcal{AP}$ rather than $S$. 

\begin{algorithm}
\caption{Parallelized search algorithm to solving the problem \textit{traceability}} \label{alg:parallel-trace}
\algorithmicrequire{ $\mathcal{B} = (t_1, t_2, \cdots, t_n)$: a blockchain of $n$ transactions; $id$: identifier of a transaction}\\
\algorithmicensure{ $\mathcal{AP}$: all the predecessors of $t_i$}
\begin{algorithmic}[1]
\Let{$\mathcal{AP}$}{$\emptyset$}
\Let{$S$}{a set with a single element $id$}
\While{$S$ is not empty}\label{alg-step:parallel-trace-3}
    \Let{$\mathcal{R}$}{\textsc{Schedule}($S$)}\label{alg-step:trace-4}
    \For{each $(id_i, \mathcal{CK}_i) \in \mathcal{R}$}\label{alg-step:parallel-trace-5}
        \State{\textit{fork thread:} $\mathcal{P}_i \gets\ $\textsc{GetPredecessors}($id_i, \mathcal{CK}_i$)}\label{alg-step:parallel-trace-6}
        \Let{$S$}{$S\setminus \{id_i\}$}
    \EndFor\label{alg-step:trace-8}
    \State{Wait until all the threads terminate}\label{alg-step:trace-9}
    \For{each $\mathcal{P}_i$ returned by the threads}\label{alg-step:trace-10}
        \For{each $id\in \mathcal{P}_i$}
            \If{$id \notin \mathcal{AP}$}
                \Let{$S$}{$S\cup \{id\}$}
                \Let{$\mathcal{AP}$}{$\mathcal{AP} \cup \{id\}$}
            \EndIf
        \EndFor
    \EndFor\label{alg-step:trace-17}
\EndWhile\label{alg-step:parallel-trace-end-while}
\State\Return{$\mathcal{AP}$}
\end{algorithmic}
\end{algorithm}

Algo.~\ref{alg:schedule} explains the particulars of generating transaction-chunk pairs, which can be parallelized from a set of transactions. It is a common bipartite graph maximum matching problem. The input is $S$, a set of transactions. The expected output is $\mathcal{R}$, the set of transaction-chunk pairs representing the queries that can be parallelized. The $\mathcal{R}$ is empty at the beginning, and $G$ is an empty bipartite graph. At line~\ref{alg-step:schedule-3}-\ref{alg-step:schedule-7}, we add vertex $v_i$ to $\mathcal{V}$ for each chunk $\mathcal{CK}_i$. Also, for each transaction belongs to $S$, we add a vertex $u_i$ to $\mathcal{U}$ representing transaction $id_i$. Next, from line~\ref{alg-step:schedule-8}-\ref{alg-step:schedule-11}, for each given transaction, we first calculate the allocated chunk index based on Algo.~\ref{alg:tx-allocation} and then add an edge to graph $G$ representing the pair of transactions and its allocated chunk index.

Until this step, we have transformed the original transaction data into the graph format in the form of transaction and its corresponding chunk index pairs, stored in $G$. Then, the problem has been modeled as a \textsc{Maximum-Matching} problem, which aims to find the maximum number of edges that share no vertex. We use the Hungarian algorithm to solve the problem. In particular, for a complete bipartite graph $G$, the Hungarian algorithm finds the maximum-weight matching, sometimes called the assignment problem. A bipartite graph can easily be represented by an adjacency matrix, where the weights of edges are the entries. The method operates on this key idea: if a number is added to or subtracted from all of the entries of any one row or column of a cost matrix, then an optimal assignment for the resulting cost matrix is also an optimal assignment for the original cost matrix. Based on this \textsc{Maximum-Matching} Algorithm, we get the result of $\mathcal{R}'$, which is the edge set of non-conflict edges. At line~\ref{alg-step:schedule-14}-\ref{alg-step:schedule-16}, we transform the returned eligible edges back into $(id_i, \mathcal{CK}_i)$ key pairs. Finally, we return the result $\mathcal{R}$ which is the input of line~\ref{alg-step:parallel-trace-5} at Algo.~\ref{alg:parallel-trace}.

\begin{algorithm}
\caption{Procedure \textsc{Schedule} as in Algo.~\ref{alg:parallel-trace} to Generate Parallelized Query} \label{alg:schedule}
\algorithmicrequire{ $S$: a set of transactions}\\
\algorithmicensure{ $\mathcal{R}$: a set of transaction-chunk pairs representing the queries that can be parallelized}
\begin{algorithmic}[1]
\Let{$\mathcal{R}$}{$\emptyset$}
\Let{$G$}{an empty bipartite graph with vertex sets $\mathcal{U}$ and $\mathcal{V}$, and edge set $\mathcal{E}$}
\For{$i \gets 0$ \textbf{to} $\alpha - 1$}\label{alg-step:schedule-3}
    \State{Add a vertex $v_i$ to $\mathcal{V}$ representing chunks $\mathcal{CK}_i$}
\EndFor\label{alg-step:schedule-5}
\For{$id_i \in S$}\label{alg-step:schedule-6}
    \State{Add a vertex $u_i$ to $\mathcal{U}$ representing transaction $id_i$}\label{alg-step:schedule-7}
    \For{$i \gets 1$ \textbf{to} $\beta$}\label{alg-step:schedule-8}
        \Let{$j$}{$id_i \bmod \alpha$}
        \State{Add an edge $(u_i, v_j)$ to $\mathcal{E}$}
    \EndFor\label{alg-step:schedule-11}
\EndFor\label{alg-step:schedule-12}
\Let{$\mathcal{R}'$}{\textsc{Maximum-Matching}$(\mathcal{G})$}
\For{each $(u_i, v_j) \in \mathcal{R}'$}\label{alg-step:schedule-14}
    \Let{$\mathcal{R}$}{$\mathcal{R} \cup \{(id_i, \mathcal{CK}_j)\}$}
\EndFor\label{alg-step:schedule-16}
\State\Return{$\mathcal{R}$}
\end{algorithmic}
\end{algorithm}

\subsection{Time Complexity Analysis}
In this subsection, we formally analyze and compare the time complexities of Algo.~\ref{alg:bfs} and Algo.~\ref{alg:parallel-trace}.

We assume the number of returned predecessors to be $m$, i.e., $\mathcal{AP} = m$. The time complexity of Algo.~\ref{alg:bfs} is $O(m\log m + mt(n))$ when maintaining the set $\mathcal{AP}$ and invoking the procedure \textsc{GetPredecessors} for $m$ times.

Because Algo.~\ref{alg:schedule} is a function called by Algo.~\ref{alg:parallel-trace}, we analyze the time complexity of Algo.~\ref{alg:schedule} first. Algo.~\ref{alg:schedule} constructs a graph and run the \textsc{Maximum-Matching} algorithm. Note that the time complexity of the \textsc{Maximum-Matching} algorithm is $O(V\cdot E)$, in which $V$ and $E$ are the numbers of vertices and edges of the graph, respectively \cite{kuhn1955hungarian}. When constructing the graph, $\alpha$ vertices are added from line \ref{alg-step:schedule-3} to \ref{alg-step:schedule-5}, and $|S|$ vertices and $|S|\cdot \beta$ edges are added from line \ref{alg-step:schedule-6} to \ref{alg-step:schedule-12}. Here, $|S| = O(m)$ because $S$ from Algo.~\ref{alg:parallel-trace} is a subset of $\mathcal{AP}$. Therefore, the number of vertices and edges are $O(\alpha + m)$ and $O(m\beta)$, respectively. To this end, the time complexity of Algo.~\ref{alg:schedule} is $O((\alpha +m)m\beta)$. Because $\alpha$ is a constant compared to $m$, the time complexity is reduced to $O(\beta\cdot m^2)$.

In Algo.~\ref{alg:parallel-trace}, we also assume that $p$ transactions are searched in parallel at line \ref{alg-step:parallel-trace-6} on average. We find that the main loop from line \ref{alg-step:parallel-trace-3} to \ref{alg-step:parallel-trace-end-while} is entered for $\frac{m}{p}$ times. Inside the main loop, line \ref{alg-step:trace-4} takes $O(\beta\cdot m^2)$ as analyzed previously and line \ref{alg-step:parallel-trace-5}-\ref{alg-step:trace-9} takes $O(p+t(n)+\log m)$ time, reduced to $O(t(n)+\log m)$ because $p$ is minor compared to $t(n)$ and $\log m$. As a result, the main loop takes $O(\frac{m}{p}\cdot (\beta\cdot m^2 +t(n)+\log m) ) = O(\frac{\beta m^3}{p} + \frac{mt(n)}{p} )$ because $\log m$ is minor compared to $\beta \cdot m^2$, excluding line \ref{alg-step:trace-10}-\ref{alg-step:trace-17}. In terms of line \ref{alg-step:trace-10}-\ref{alg-step:trace-17}, it takes $O(m\log m)$ in total because its purpose is to maintain two sets $S$ and $\mathcal{AP}$, both of which are of size $O(m)$. As a result, the overall time complexity of Algo.~\ref{alg:parallel-trace} is $O(m\log m + \frac{\beta m^3}{p} + \frac{mt(n)}{p} ) = O(\frac{\beta m^3}{p} + \frac{mt(n)}{p})$ because $m\log m$ is minor compared to $\beta m^3 / p$.

Next, we compare the time complexities of Algo.~\ref{alg:bfs} and Algo.~\ref{alg:parallel-trace}, which are $O(m\log m + mt(n))$ and $O(\frac{\beta m^3}{p} + \frac{mt(n)}{p})$, respectively. If $t(n)$ dominates the time complexity compared to $m$ (for example, when $n$ is large), Algo.~\ref{alg:parallel-trace} will take much less time than Algo.~\ref{alg:bfs} theoretically. This is because $t(n)$ is averaged by $p$ times in Algo.~\ref{alg:parallel-trace}. Note that Algo.~\ref{alg:parallel-trace} achieves much lower time overhead with the sacrifice in higher storage overhead ($\beta - 1$ replicas of the transactions in $\alpha$ trunks).

\section{Experimental Results \& Discussion}\label{sec:experiments}
In this section, we demonstrate the effectiveness and practicability of the proposed high-efficiency traceability solution based on implementation on Hyperledger Fabric \cite{androulaki2018hyperledger} and extensive experiments evaluating the parallelization ratio and storage ratio. We also discuss the transaction allocation algorithm and the database selection, which might affect the tracing efficiency in this work.

\subsection{Experimental Environments \& Design}
\begin{figure}[!ht]
    \centering
    \includegraphics[width=\linewidth]{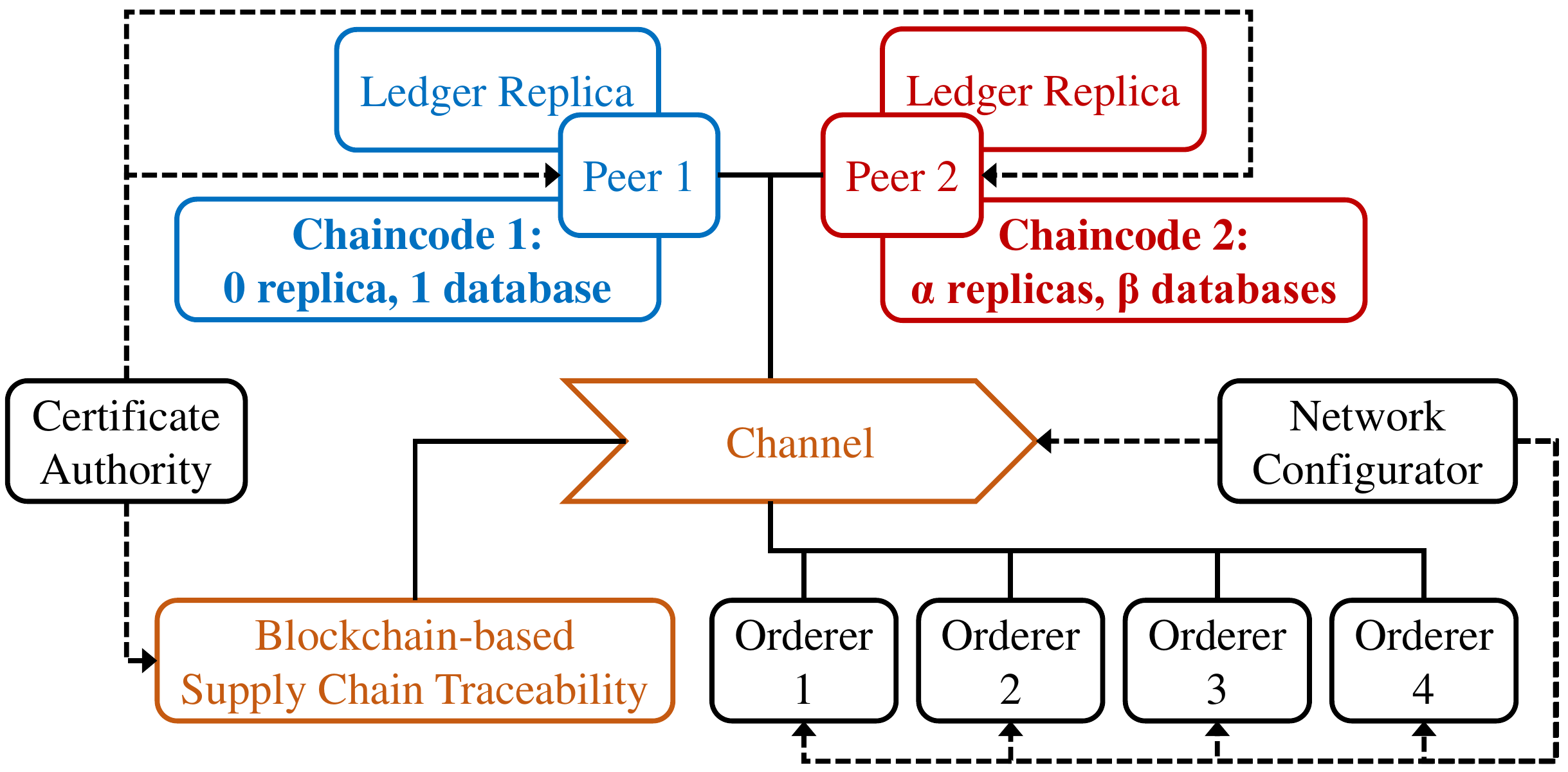}
    \caption{System architecture. The blockchain-based supply chain traceability system and algorithms are implemented in Hyperledger Fabric with $4$ orderers, $2$ peers, $2$ chaincodes, and $1$ channel.}\label{fig:deployment}
\end{figure}

In this work, we leverage Hyperledger Fabric, an open-source permissioned blockchain platform, to implement and evaluate our proposed algorithms. Fig.~\ref{fig:deployment} depicts the architecture of the developed system using Hyperledger Fabric, showing the components and their relationship as follows:
\begin{itemize}
\item The benchmark algorithm, i.e., BFS-based solution when $\alpha = 0$ and $\beta = 1$, is implemented in ``Chaincode $1$'', hosted by ``Peer $1$''.
\item The proposed algorithm in this work with varying parameters $\alpha$ and $\beta$ is implemented in ``Chaincode $2$'', hosted by ``Peer $2$''.
\item The two chain codes, i.e., ``Chaincode $1$'' and ``Chaincode $2$'', are deployed on the ``Channel'', supporting the application ``Blockchain-based Supply Chain Traceability''.
\item The transactions in ``Channel'' are ordered by four orderers, i.e., ``Orderer $1$'', ``Orderer $2$'', ``Orderer $3$'', and ``Orderer $4$'', running the crash fault-tolerant consensus protocol as provided by Hyperledger Fabric.
\item The ``Certificate Authority'' dispenses identities to the application and two peers.
\item The ``Network Configurator'' configures the networks of the channel and four orderers.
\end{itemize}

\begin{figure}[!ht]
    \centering
    \includegraphics[width=\linewidth]{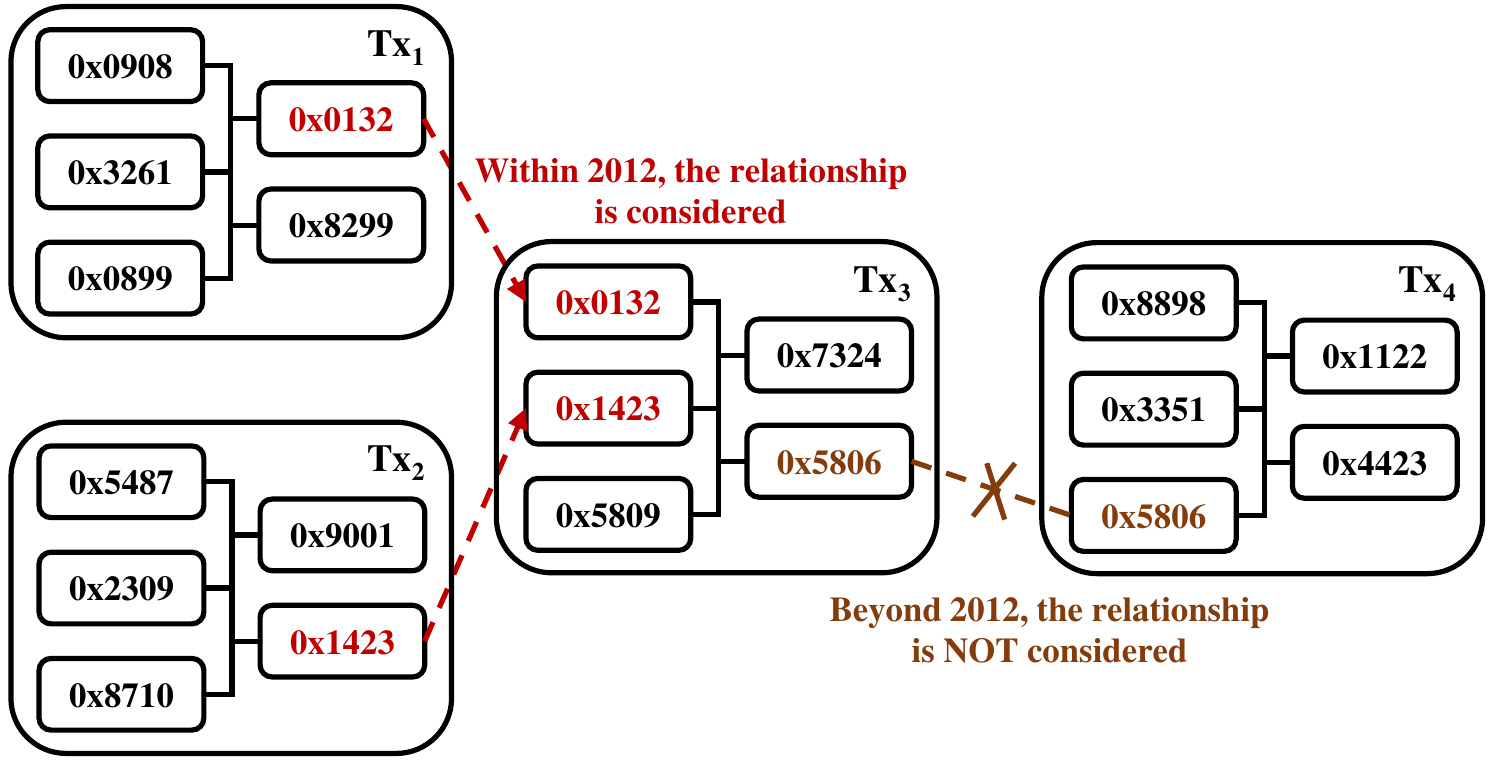}
    \caption{An example of how Bitcoin transactions are used in the experiments. Particularly, $Tx_1$, $Tx_2$, and $Tx_3$ are within $2012$ and considered supply chain transactions while $Tx_4$ is not. Because $0x0132$ is the output of $Tx_1$ and the input of $Tx_3$, we notate $Tx_1$ as one of the \textit{direct predecessors} of $Tx_3$. The relationship between $Tx_3$ and $Tx_4$ is not considered because $Tx_4$ is beyond $2012$.}\label{fig:bit-example}
\end{figure}

We deploy a prototype based on the system architecture using eight workstations. Each workstation runs Ubuntu $20.04$, consisting of a $4$-core $8$-thread Intel Core i7-8809G $4.2$Ghz CPU, a $32$GB DDR4 DRAM, and a $1,024$GB NVMe SSD. The workstations are connected in a local network, forming a blockchain network. The implementation and deployment imply the practicability of the proposed high-efficiency traceability solution. We use the Bitcoin data in $2012$, containing up to $1.9$ million transactions, as the input of the blockchain-based traceability system. In particular, each Bitcoin transaction $tx$ is regarded as a supply chain transaction, containing several inputs ($\{in_1, in_2, \cdots\}$) and outputs ($\{out_1, out_2, \cdots\}$). For each $in_i$, it must be the output of another transaction $tx'$ because of the safety of the Bitcoin system. If $tx'$ is also generated in $2012$, we notate $tx'$ as one of the \textit{direct predecessors} of $tx$. Similarly, if $out_j$ is the input of another transaction $tx''$ and $tx''$ is generated in $2012$, we notate $tx$ as one of the \textit{direct predecessors} of $tx''$. Fig.~\ref{fig:bit-example} depicts an example of the data usage.

We study how the proposed solution performs with the prototype system compared with the BFS-based solution. Based on Algo.~\ref{alg:tx-allocation}, Algo.~\ref{alg:parallel-trace}, and Algo.~\ref{alg:schedule}, three variables will impact the proposed solution's efficiency: ${n}$, $\alpha$, and $\beta$, representing the numbers of transactions, chunks, and replicas, respectively.

In order to reduce the impact of $n$ on the system, we use \textit{parallelization ratio} and \textit{storage ratio} as two key performance metrics for demonstrating the solution's effectiveness. The \textit{parallelization ratio} and \textit{storage ratio} are the execution time overhead and chunk storage overhead compared to the BFS-based solution, respectively. In the following experiments, we will examine the parallelization and storage ratios with combinations of $\alpha$ and $\beta$. We will also discuss the transaction allocation algorithm and database selection. We repeat the experiment $50$ times for each experiment to get the average results.

\subsection{Evaluation of Parallelization Ratio}\label{evaofpra}
In this experiment, we will compare the parallelization ratio of the proposed parallelization algorithm. We try to find the pair of optimal parameters of $\alpha$ and $\beta$, by changing their values, calculated by the number of operations. The algorithm has such a condition that it will be terminated by final results where the transaction has no father nodes or is the genesis/origin transaction. Intuitively, with the increase of $\alpha$ (the number of chunks) or $\beta$ (the number of replicas), the parallelization ratio shall also rise compared to the straightforward BFS-based solution.

\begin{figure}[!ht]
    \centering
    \includegraphics[width=.7\linewidth]{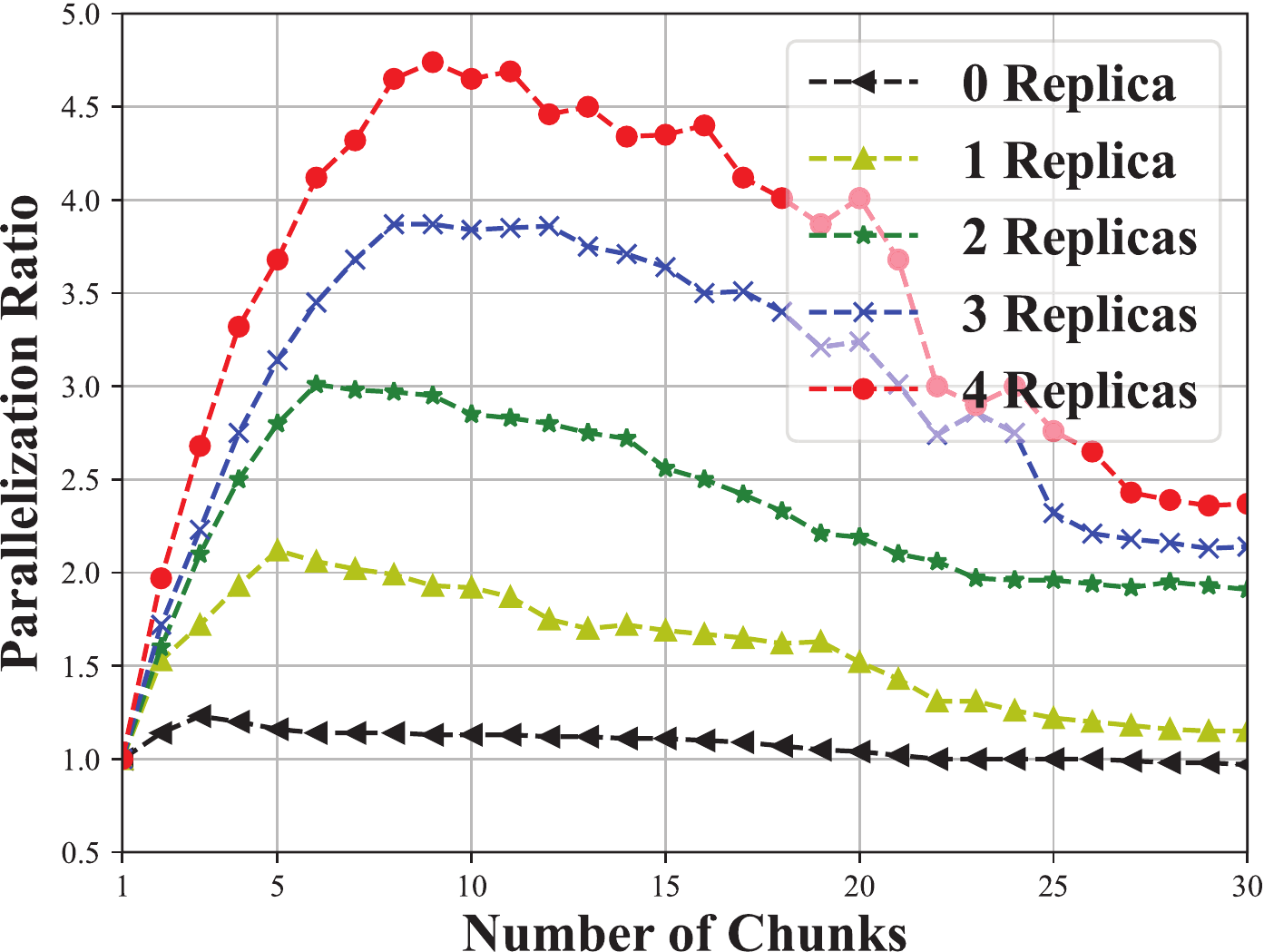}
    \caption{Parallelization ratio with $0-4$ replicas. Despite the number of replicas, the parallelization ratio will increase first and then decrease with the increasing number of chunks. The number of replicas ($0-4$) significantly affects the parallelization ratio.}\label{fig:ex-1-5}
\end{figure}

\begin{figure}[!ht]
\centering
\includegraphics[width=.7\linewidth]{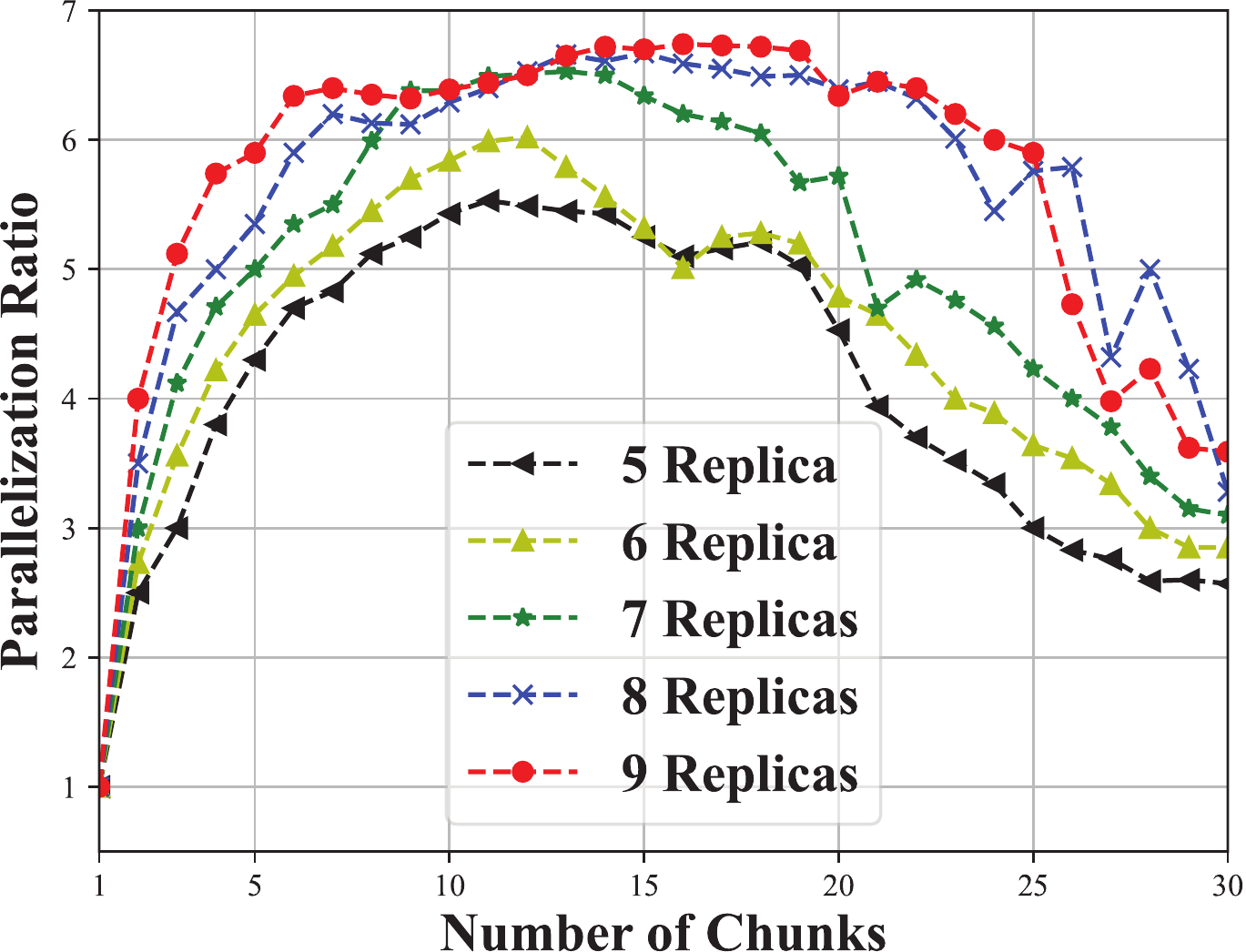}
\caption{Parallelization ratio with 5-9 replicas. The influence of the number of replicas on the parallelization ratio is less and less noticeable with an increasing number of replicas.}\label{fig:ex-6-10}
\end{figure}

\begin{figure}[!ht]
\centering
\includegraphics[width=.7\linewidth]{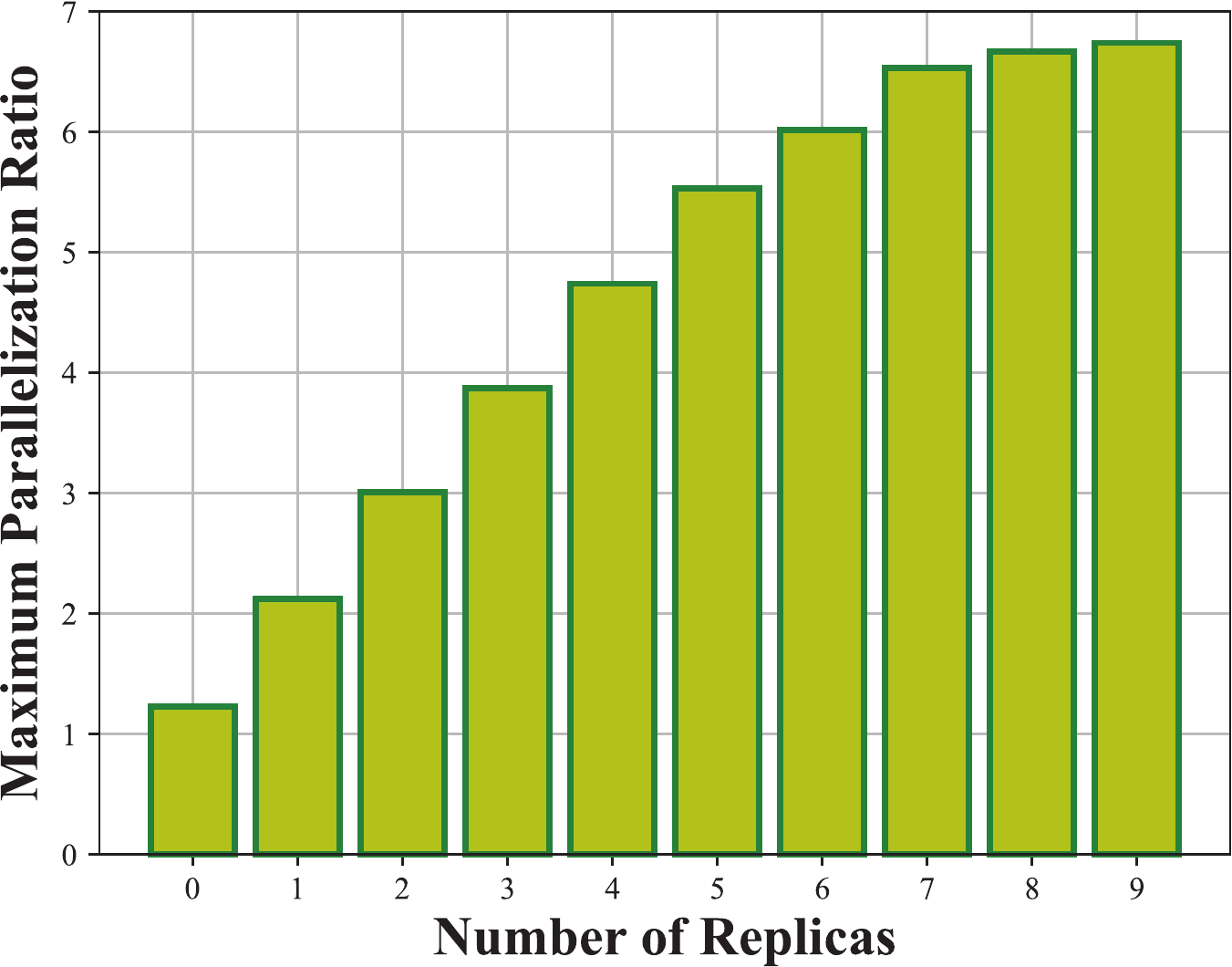}
\caption{Maximum parallelization ratio that can be achieved with different numbers of replicas.}\label{fig:ex-max-ratio}
\end{figure}

Fig.~\ref{fig:ex-1-5} and Fig.~\ref{fig:ex-6-10} depict the change of parallelization ratio with $1$ to $4$ replicas and $5$ to $9$ replicas respectively. Note that the $0$ replica and $1$ chunk indicate the BFS-based solution. It is obvious that when the number of chunks is fixed, the parallelization ratio increases dramatically with the increase of replicas. When there are $9$ replicas, the parallelization ratio can be up to $6.74$ as shown in Fig.~\ref{fig:ex-6-10} and Fig.~\ref{fig:ex-max-ratio}, which means $1 - \frac{1}{6.74} \approx 85.1\%$ time can be saved.

Such a trend holds when the number of chunks is small. More precisely, the turning point slightly shifts towards the right as the number of replicas increases. For example, when $\beta$ equals $2$, the turning point of $\alpha$ is around $6$. When $\beta$ equals $6$, the turning point of $\alpha$ is around $12$. For the surge part, the reason is that each transaction has a higher probability of coverage to store it into different chunks when the number of replicas increases. After the turning point, the curve comes down moderately. The reason is that the increasing number of chunks decreases the probability of finding the exact transaction among chunks. In extreme cases, the ratio becomes unstable at the end when having $8$ and $9$ replicas. The reason is that the relatively large number of replicas distributed among chunks increases the complexity of finding the target transaction. Fig.~\ref{fig:ex-max-ratio} depicts the maximum parallelization ratio that can be achieved based on the different number of replicas. The growth slows down when the system has more than six replicas. 

\begin{figure}[!ht]
\centering
\includegraphics[width=.7\linewidth]{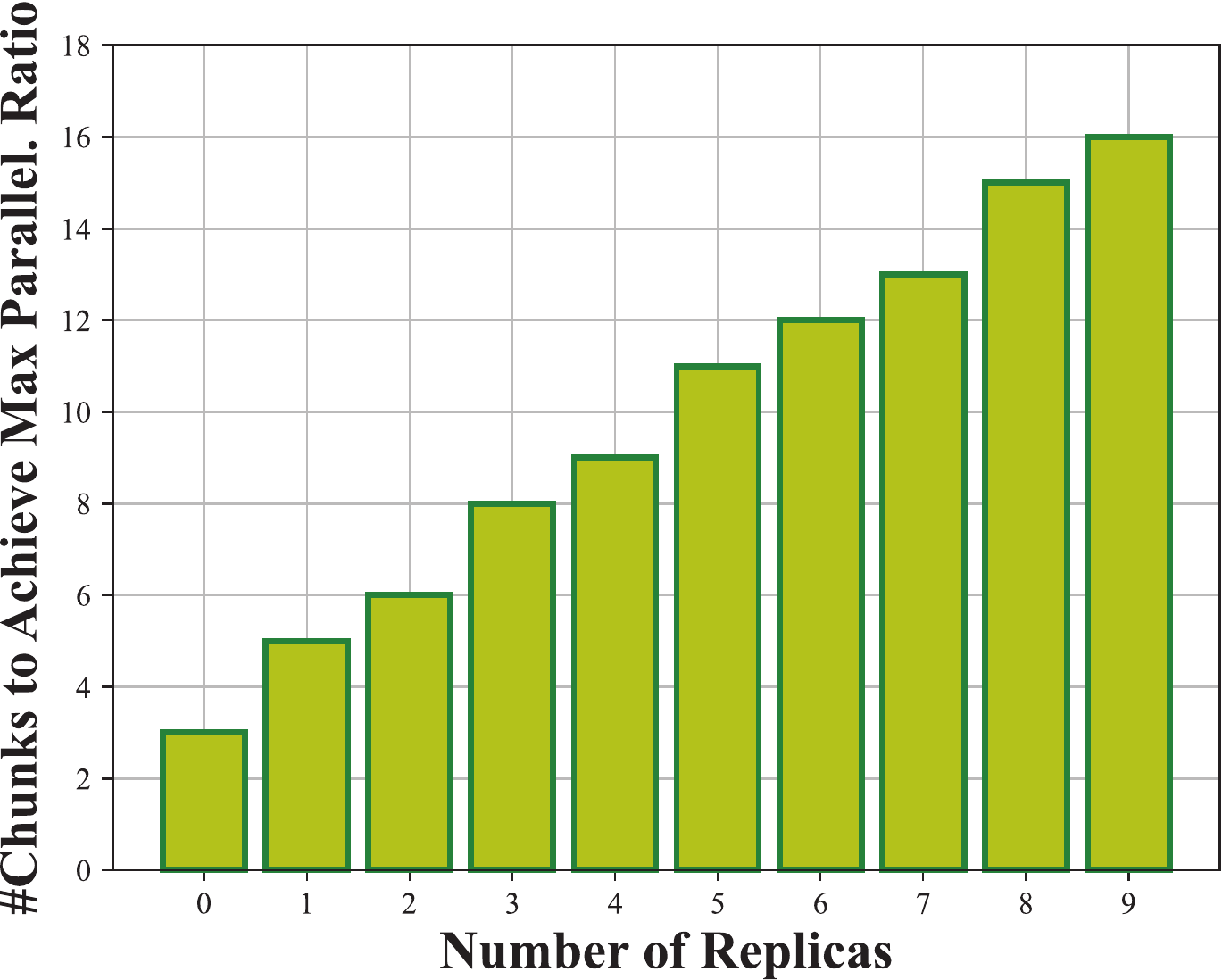}
\caption{Number of chunks to achieve the maximum parallelization ratio. It requires more chunks to achieve the maximum parallelization ratio as the number of replicas increases.}\label{fig:ex-max-chunk}
\end{figure}

In Sec.~\ref{evaofpra}, we plan to find the optimal pair of $\alpha$ and $\beta$ via changing their values. For each $\beta$, there should be an optimal value of $\alpha$ which should be noticed at the curve's apex. Fig.~\ref{fig:ex-max-chunk} illustrates the relationship between the number of chunks to achieve the max parallelization ratio with the number of replicas. It is not hard to find that such a linear relationship maintains steadily with more replicas. By simple linear regression, we can get $f(\alpha)=1.43\beta+1.93$. This formula implies that the number of chunks should not be too small or too large, given the number of replicas. The excess number of chunks does not contribute to the parallelization ratio, which has an upper limit, as shown in Fig.~\ref{fig:ex-max-ratio}. The ``sweet point" of the $\alpha$ and $\beta$ can be easily calculated, which will be helpful when we have more replicas.

\subsection{Evaluation of Storage Ratio}
In this set of experiments, we investigate the database storage overhead of the proposed parallelization algorithm. We will fix the value of $\alpha$ and change the value of $\beta$. Then we do it in a reverse way by fixing the $\beta$ and changing the $\alpha$. The storage cost of the BFS-based solution is normalized as $1$ for easier comparison.

\begin{figure}[!ht]
  \centering
  \includegraphics[width=.7\linewidth]{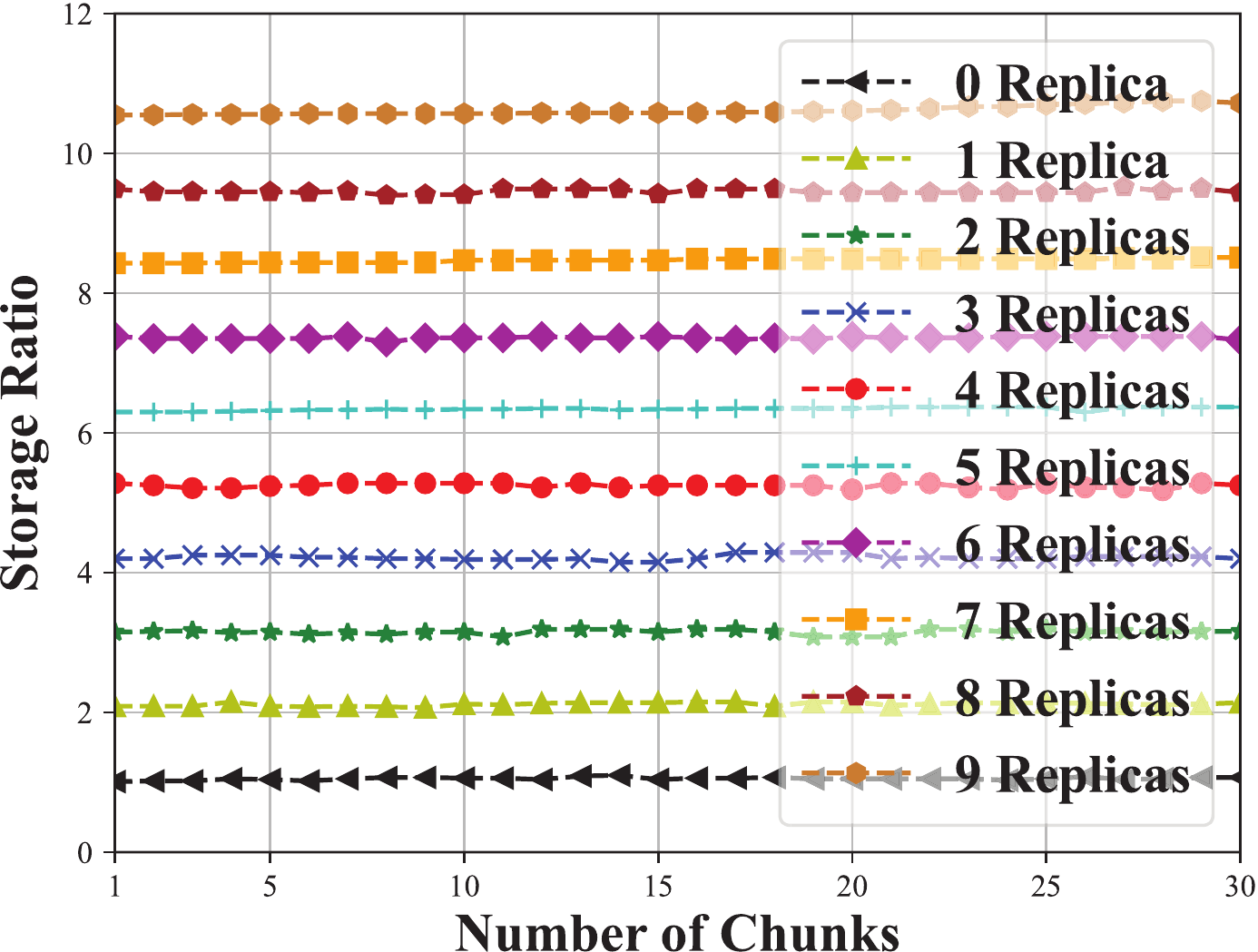}
  \caption{Database storage ratio with $0-9$ replicas. The storage ratio remains nearly unchanged with different numbers of chunks.}
  \label{fig:storratio-0-9}
\end{figure}

Fig.~\ref{fig:storratio-0-9} illustrates the change of storage overhead ratio with different settings of the number of chunks and replicas. The experimental results show that the storage overhead is largely affected by the value $\beta$, the number of replicas. With more replicas available, the storage ratio grows steadily. On the other hand, $\alpha$, i.e., the number of chunks, affects little on the storage ratio, with a slight increase when more chunks are used. The reason for the slight increase is that more chunks lead to higher storage overhead in building the chunk index. Overall, the storage overhead slightly increases compared with the BFS-based solution, which is acceptable to our concern.

\subsection{Discussion of Transaction Allocation Algorithm}
In Sec.~\ref{subsec:prosol}, we point out that the major drawback of the traditional approach is the frequent operations of \textsc{GetPredecessors} with high time overheads. Based on the insight, we propose using $\alpha$ chunks to store the transactions, which can dramatically improve the parallelized operations when tracing the transactions. We propose a straightforward transaction allocation mechanism, Algo.~\ref{alg:tx-allocation}, with the $\bmod$ operation. However, the issue of imbalance may still occur for many reasons. For example, the uneven distribution of transaction identifiers can be found in many statistical reports, which may be solved using the $\bmod$ function.

To improve the randomness of transaction allocation, we propose a new random mechanism as shown in Algo.~\ref{alg:tx-allocation-random}. The main difference with the Algo.~\ref{alg:tx-allocation} is that before $\bmod$ each transaction identifier with $\alpha$, Algo.~\ref{alg:tx-allocation-random} pre-processes the transaction identifier with a hash function $f_i$ (e.g., SHA and MD5). To this end, the deterministic correlation between the transaction identifier and the target allocated chunk can be eliminated. It is expected that the randomness can contribute to a more average distribution of the transactions on the chunks and lower the tracing time overhead in consequence.

\begin{algorithm}
\caption{Transaction allocation (random method)} \label{alg:tx-allocation-random}
\algorithmicrequire{ $id_i$: identifier of the new transaction; $\mathcal{P}_i$: the set of direct predecessors of the new transaction}\\
\algorithmicensure{ The allocation scheme of the new transaction}
\begin{algorithmic}[1]
\For{$i \gets 1$ \textbf{to} $\beta$}
    \State{Store $(id_i, \mathcal{P}_i)$ in $\mathcal{CK}_{f_i(id_i) \bmod{\alpha}}$}
\EndFor
\end{algorithmic}
\end{algorithm}

We conduct experiments to compare Algo.~\ref{alg:tx-allocation} and Algo.~\ref{alg:tx-allocation-random}. We find that the two algorithms perform nearly the same in parallelization and storage ratios. In particular, the two algorithms lead to similar distributions regarding vertex degree. It means Algo.~\ref{alg:tx-allocation-random} can hardly achieve better randomness of transaction allocation than Algo.~\ref{alg:tx-allocation} as expected.

\subsection{Discussion of Database Selection}
Blockchain can be considered a multi-node database maintained by a network of independent participants. It is decentralized, with no single user having the ultimate authority over the system. On the other hand, the database, unlike blockchains, are a centralized ledger that is run by an administrator. Although blockchain looks contradictory to the database, they are closely connected. Conceptually, as a whole, a blockchain is distributed across the entire network of peers. Fundamentally, a single network node still relies on a specific database to maintain its local ledger, synchronized with peers, for verification and synchronization purposes. For example, the Bitcoin core client uses the LevelDB database for the block index and the chain state, also known as the unspent transaction output set. LevelDB is also the default key-value state database embedded in Hyperledger Fabric. At the same time, CouchDB is a choice as it supports rich queries and indexing for more efficient queries over large datasets. During our experiments,  we find that the database selection does not impact the parallelization ratio.

\section{Conclusion and Future Directions}\label{sec:conclusion}
This work is the first to study the efficiency issue of blockchain-based supply chain traceability. First, we depict the system model supply chain and formally define the traceability problem as a graph searching problem. Then, a parallel searching algorithm is proposed, in which the maximum flow theory is employed and adapted to maximize the parallelization ratio. The experimental results show up to $85.1$\% reduction in the product tracking time. The proposed algorithm is expected to be applied for broader applications, e.g., tracking of cryptocurrencies, besides supply chain traceability.

In the future, we will study the algorithm to further boost the time efficiency of blockchain-based supply chain traceability. Particularly, the proposed algorithm contains a sequence of allocations of search queries. This work decides the allocations one by one without considering the influence of the current allocation on the future ones. We will consider the sequence of search allocations to reduce the tracing time overhead. Moreover, The experiments in this work are based on Bitcoin data because real-world supply chain data can hardly be obtained. We will validate the proposed algorithm on real-world supply chains given available data.

\section{Acknowledgment}
This work is supported by the Research Institute for Artificial Intelligence of Things, The Hong Kong Polytechnic University, HK RGC CRF No. C2004-21GF, and HK RGC RIF Noo. R5034-18.

\bibliographystyle{IEEETran}
\bibliography{refs5}
\begin{IEEEbiography}[{\includegraphics[width=1in,height=1.25in]{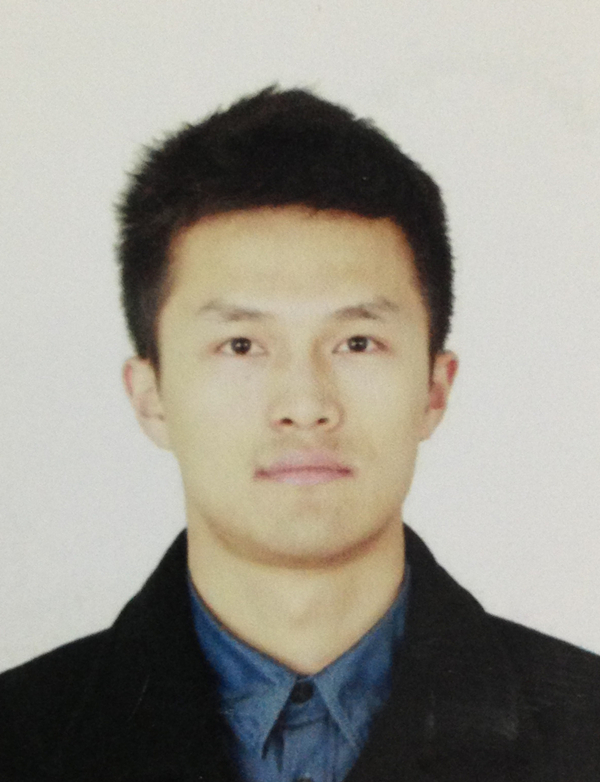}}]{Hanqing Wu}
is working towards the Ph.D. degree in computer science with the Department of Computing, The Hong Kong Polytechnic University, Hong Kong SAR, China. Before that, he was a research assistant with The Hong Kong Polytechnic University from May 2015 to December 2015. He received the B.Sc. degree in software engineering from Tongji University in 2010. His research interests include distributed computing, blockchain, and big data.
\end{IEEEbiography}

\begin{IEEEbiography}[{\includegraphics[width=1in,height=1.25in]{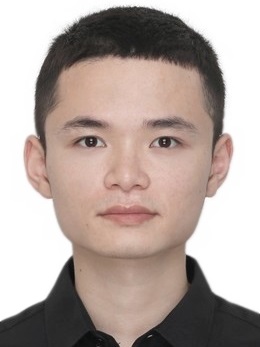}}]{Shan Jiang}
received the B.Sc. degree in computer science and technology from Sun Yat-sen University, Guangzhou, China, in 2015 and the Ph.D. degree in computer science from The Hong Kong Polytechnic University, Hong Kong SAR, in 2021. He is currently a Research Assistant Professor with the Department of Computing, The Hong Kong Polytechnic University, Hong Kong SAR. Before that, he visited Imperial College London from November 2018 to March 2019. He won the best paper award from BlockSys 2021 International Conference on Blockchain and Trustworthy Systems. His research interests include distributed systems and blockchain, blockchain-based big data sharing, and blockchain as a service.
\end{IEEEbiography}

\begin{IEEEbiography}[{\includegraphics[width=1in,height=1.25in,clip,keepaspectratio]{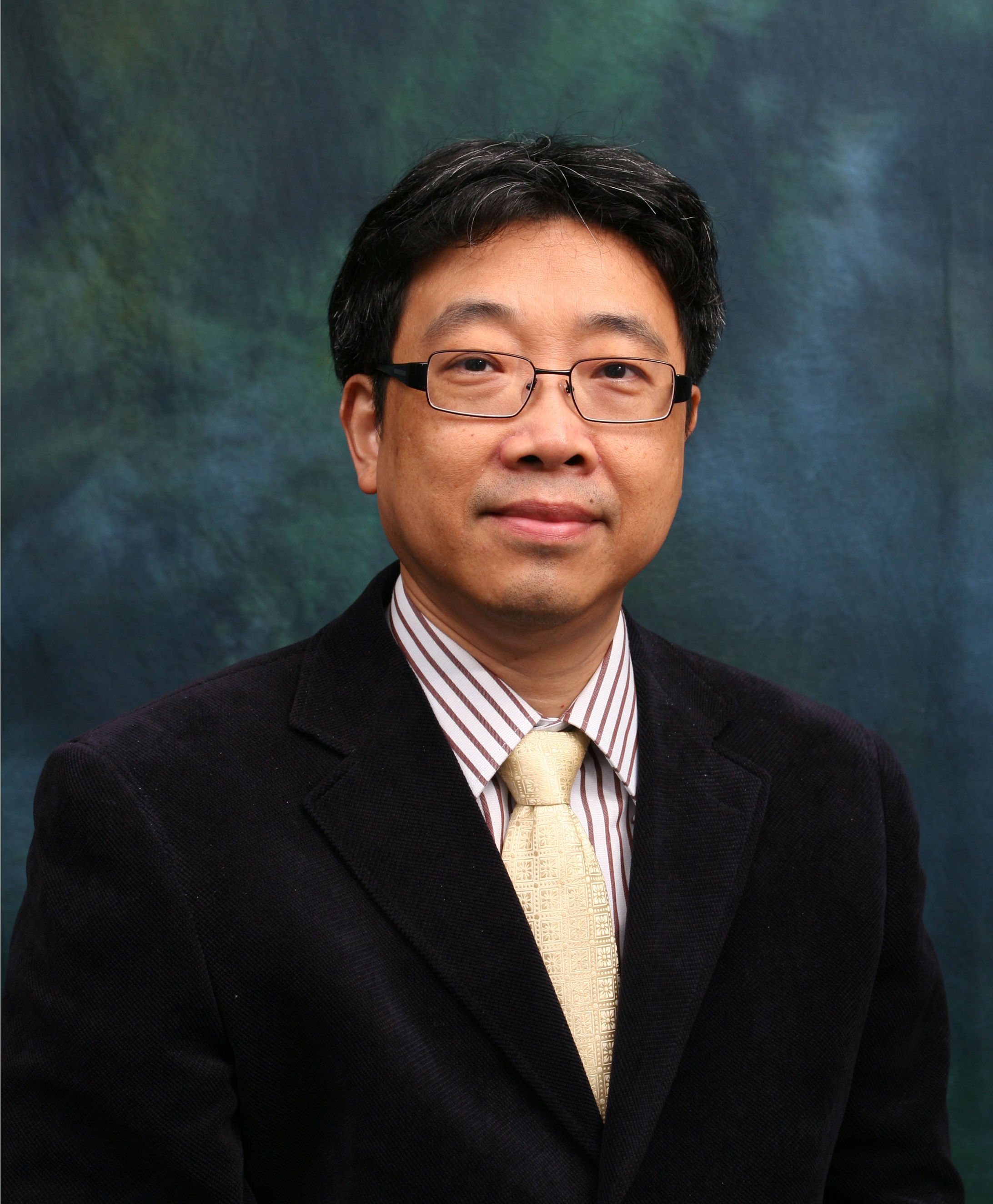}}]{Jiannong Cao}
(M'93-SM'05-F'15) received the B.Sc. degree in computer science from Nanjing University, Nanjing, China, in 1982, and the M.Sc. and Ph.D. degrees in computer science from Washington State University, WA, USA, in 1986 and 1990, respectively.
He is currently the Otto Poon Charitable Foundation Professor in Data Science and the Chair Professor of Distributed and Mobile Computing in the Department of Computing at The Hong Kong Polytechnic University (PolyU), Hong Kong. He is also the Dean of Graduate School, the director of the Research Institute for Artificial Intelligence of Things in PolyU, and the director of the Internet and Mobile Computing Lab. He was the founding director and is now the associate director of PolyU's University Research Facility in Big Data Analytics. He served the department head from 2011 to 2017. Prof. Cao is a member of Academia Europaea, a fellow of IEEE, a fellow of the China Computer Federation (CCF), and an ACM distinguished member. His research interests include distributed systems and blockchain, wireless sensing and networking, big data and machine learning, and mobile cloud and edge computing.
\end{IEEEbiography}

\end{document}